\documentclass[english,aps,prl,superscriptaddress,twocolumn]{revtex4}

\usepackage[T1]{fontenc}

\usepackage{graphicx}
\usepackage{epstopdf}
\usepackage{amssymb}
\usepackage{amsmath}
\usepackage{subfigure}
\usepackage[urlcolor=blue,hyperindex,colorlinks,bookmarks=true,linkcolor=black,citecolor=black]{hyperref}
\usepackage[normalem]{ulem}
\usepackage[abs]{overpic}
\usepackage{color}
\usepackage{rotating}

 \newcommand{\be}{\begin{equation}}
\newcommand{\ee}{\end{equation}}
\newcommand{\bea}{\begin{align}}
\newcommand{\eea}{\end{align}}

\newcommand{\ket}[1]{\left|#1\right\rangle}
\newcommand{\bra}[1]{\left\langle#1\right|}
\newcommand{\braket}[2]{\left\langle#1\right|\left.#2\right\rangle}

\newcommand{\abs}[1]{\lvert#1\rvert}

\newcommand{\yy}{\gamma}
\newcommand{\al}{\alpha}
\newcommand{\ww}{\omega}
\newcommand{\eps}{\epsilon}

\begin{document}

\title{Scalable two- and four-qubit parity measurement with a threshold photon counter}

\author{Luke C.G. Govia}
\email[Electronic address: ]{lcggovia@lusi.uni-sb.de}
\affiliation{Theoretical Physics, Universit\"{a}t des Saarlandes, Saarbr\"{u}cken, Germany}

\author{Emily J. Pritchett}
\altaffiliation[Present affiliation: ]{HRL Laboratories, LLC, Malibu, CA 90265, USA}
\affiliation{Theoretical Physics, Universit\"{a}t des Saarlandes, Saarbr\"{u}cken, Germany}

\author{B.L.T. Plourde}
\affiliation{Department of Physics, Syracuse University, Syracuse, NY 13244-1130, USA}

\author{Maxim G. Vavilov}
\affiliation{Department of Physics, University of Wisconsin, Madison, WI 53706, USA}

\author{R. McDermott}
\affiliation{Department of Physics, University of Wisconsin, Madison, WI 53706, USA}

\author{Frank K. Wilhelm}
\affiliation{Theoretical Physics, Universit\"{a}t des Saarlandes, Saarbr\"{u}cken, Germany}

\begin{abstract}
Parity measurement is a central tool to many quantum information processing tasks. In this Letter, we propose a method to directly measure two- and four-qubit parity with low overhead in hard- and software, while remaining robust to experimental imperfections. Our scheme relies on dispersive qubit-cavity coupling and photon counting that is sensitive only to intensity; both ingredients are widely realized in many different quantum computing modalities. For a leading technology in quantum computing, superconducting integrated circuits, we analyze the measurement contrast and the back action of the scheme and show that this measurement comes close enough to an ideal parity measurement to be applicable to quantum error correction. 
\end{abstract}

\maketitle
The ability to measure a quantum system in a high fidelity and quantum non-demolition (QND) way is fundamental to most aspects of quantum information processing (QIP). In the circuit quantum electrodynamics (cQED) community, great success has been achieved in qubit readout by linear amplification and homodyne detection of the signal in a dispersively coupled microwave resonator (cavity) \cite{Blais2004,Gambetta:2007qf,Jeffrey:2014zr,Vijay2011}. However, the macroscopic size and complicated circuitry required for this readout scheme is a major obstacle to scalability. For this reason a simpler, scalable, high-fidelity, QND single-qubit readout scheme based on threshold photodetection was recently introduced \cite{Govia:2014fk}.

In addition, QND readout of multi-qubit operators is increasingly important in contemporary QIP. In particular, readout of the parity of multiple qubits has applications to quantum error correction \cite{Nielsen00} such as the surface code \cite{Fowler:2012sf}, to quantum phase estimation \cite{Peruzzo:2014fj}, to the implementation of multi-qubit gates \cite{Gottesman:1999fk,Beenakker:2004kq}, and to entanglement generation \cite{Hutchison:2009qy,Riste2013,Ruskov:2003jk}. Parity measurement of two superconducting qubits has been demonstrated \cite{Riste2013,Chow:2014fk}, as has Bell-state measurement \cite{Steffen2013} and parity measurement of a cavity state using a superconducting qubit \cite{Vlastakis2013}.  However, adapting these protocols to more than two qubits requires a significant increase in circuit complexity, either in the number of microwave resonators necessary for amplification-based direct parity measurement \cite{DiVincenzo2013,Tornberg:2014nr}, or in the number of costly entangling gates in gate-based parity measurement.

In this Letter, we propose a protocol for QND readout of the parity of multiple qubits coupled to one resonator, over a timescale comparable to that of a single entangling gate, following Ref.~\cite{Govia:2014fk}. We describe the physical model, briefly review Ref.~\cite{Govia:2014fk}, and show how the procedure generalizes to an $N$-qubit single-shot parity measurement. Finally, we analyze the main sources of error. 

We describe parity measurement on $N$ qubits, each coupled to the same single-mode cavity, which is also coupled to a tunable photon counter. Each qubit ${\rm Q}_n$ of transition frequency $\omega_{{\rm Q}_n}$ couples to the cavity via a Jaynes-Cummings interaction in the dispersive regime, leading to the full cavity-qubit Hamiltonian ($\hbar=1$)
\begin{equation}
H=   H_{\rm D}+\omega_{\rm C}\hat{a}^\dagger\hat{a} +\sum_n^N \left( \chi_{{\rm Q}_n}\hat{a}^\dagger\hat{a} - \frac{\omega_{{\rm Q}_n}-\chi_{{\rm Q}_n}}{2}\hat{\sigma}_z^n\right), 
\end{equation}
where $\chi_{{\rm Q}_n} \equiv {g_{{\rm Q}_n}^2}/(\ww_{\rm C}-\ww_{{\rm Q}_n})$ is the dispersive shift for coupling strength $g_{{\rm Q}_n}$, and $H_{\rm D}\equiv  A(t)\left( \hat{a}+ \hat{a}^{\dagger}\right) $ is a time-dependent classical drive that controls the cavity. For this Hamiltonian the computational basis states are defined as the eigenvectors of $\sum_n^N \hat{\sigma}_z^n$. We assume that the qubits are far enough detuned that cavity-mediated qubit-qubit coupling is negligible. This model is sufficiently general to encompass many qubit architectures \cite{Devoret13, Jeffrey:2014zr,Riste2013,Mallet09,DiVincenzo2013,Blais2004}.

Our protocol can be applied to any physical system involving a threshold photon counter and strong dispersive coupling between qubits and electromagnetic radiation. In the microwave regime, where photon counting is possible with a Josephson Photomultiplier (JPM) \cite{Chen2011}, other examples include NV centers in diamond \cite{Kubo:2010fk}, Rydberg atoms \cite{Pritchard:2014qf,Hogan:2012xy}, and lateral quantum dots \cite{Frey:2012nr}. At higher frequencies, where photon counters are commonplace, candidate systems include trapped atoms \cite{Miller:2005zl} and self-assembled quantum dots \cite{Miguel-Sanchez:2013qf}.

As described in Ref.~\cite{Govia:2014fk}, the cavity-mediated QND measurement of a single qubit with a photon counter is performed in a three-stage protocol. First the cavity, initialized to vacuum, is driven at its shifted resonance frequency for the qubit in its excited state, populating the cavity conditionally upon the state of the qubit. In the following measurement stage, the photon counter distinguishes between the two conditional cavity states (one of which is above and one below its threshold), and therefore detects the state of the qubit. Finally, an additional microwave drive resets the cavity approximately to the vacuum state for further computation. Two properties of the counter are crucial: it should be a square-law detector that responds to the total energy only (insensitive to phase), and it should have threshold behavior at a suitable number of photons that is well between vacuum and the selectively excited state of the cavity. Number resolution and single-photon sensitivity are {\em not} required.
\begin{figure}[h!]
\subfigure{
\label{fig:Circuit}
\includegraphics[width = \columnwidth]{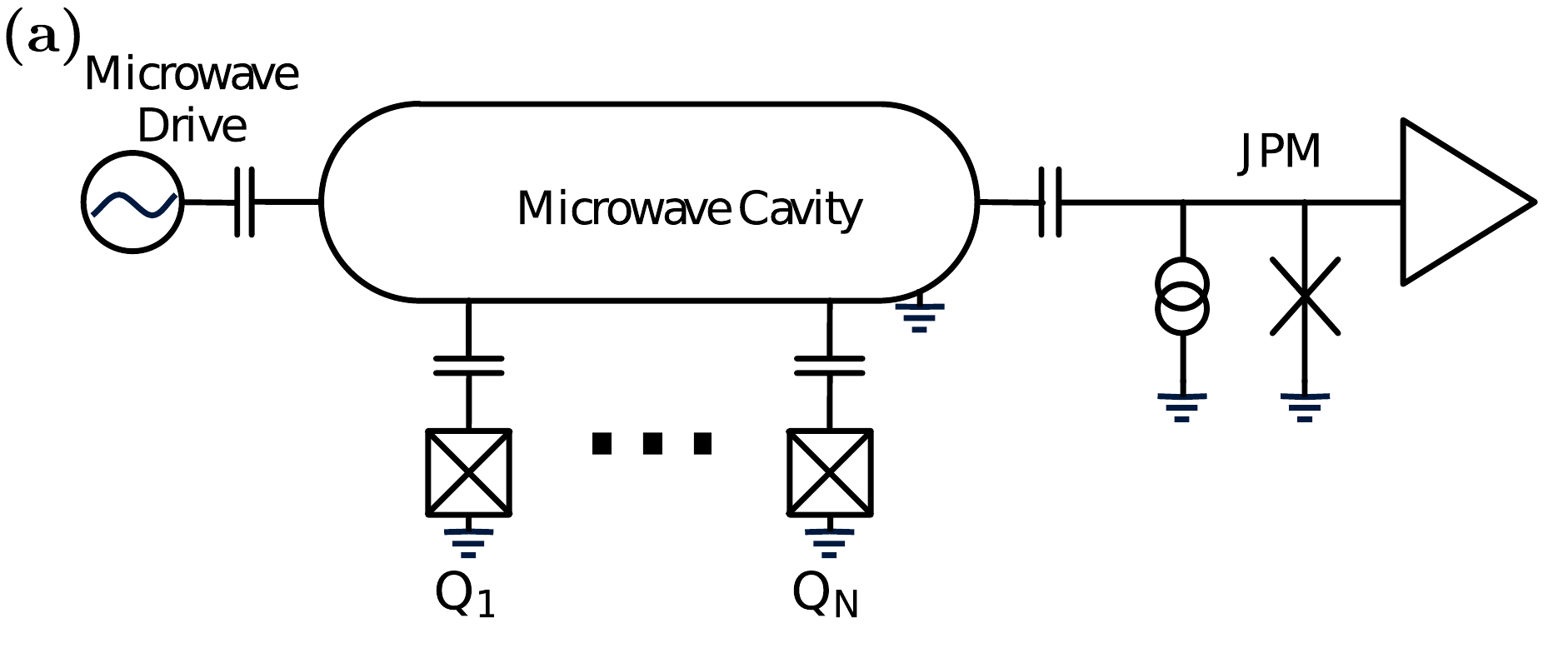}}
\subfigure{
\label{fig:2PBands}
\includegraphics[width = \columnwidth]{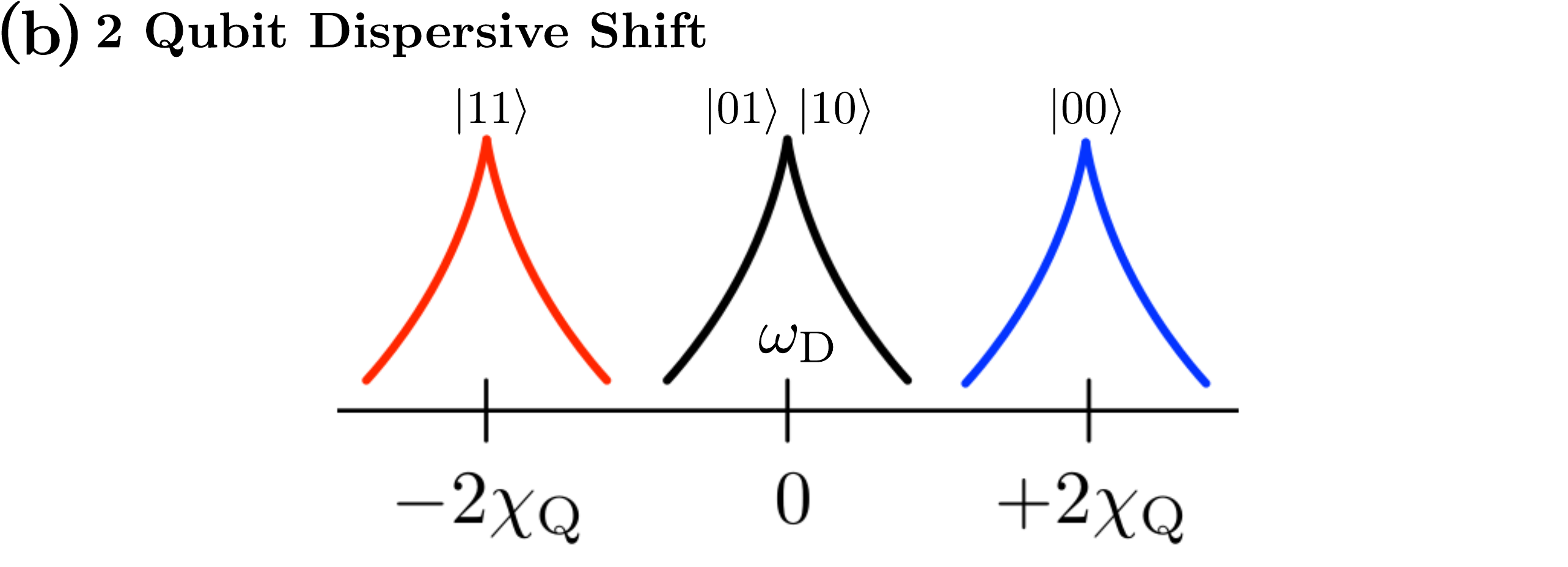}}
\subfigure{
\label{fig:PBands}
\includegraphics[width = \columnwidth]{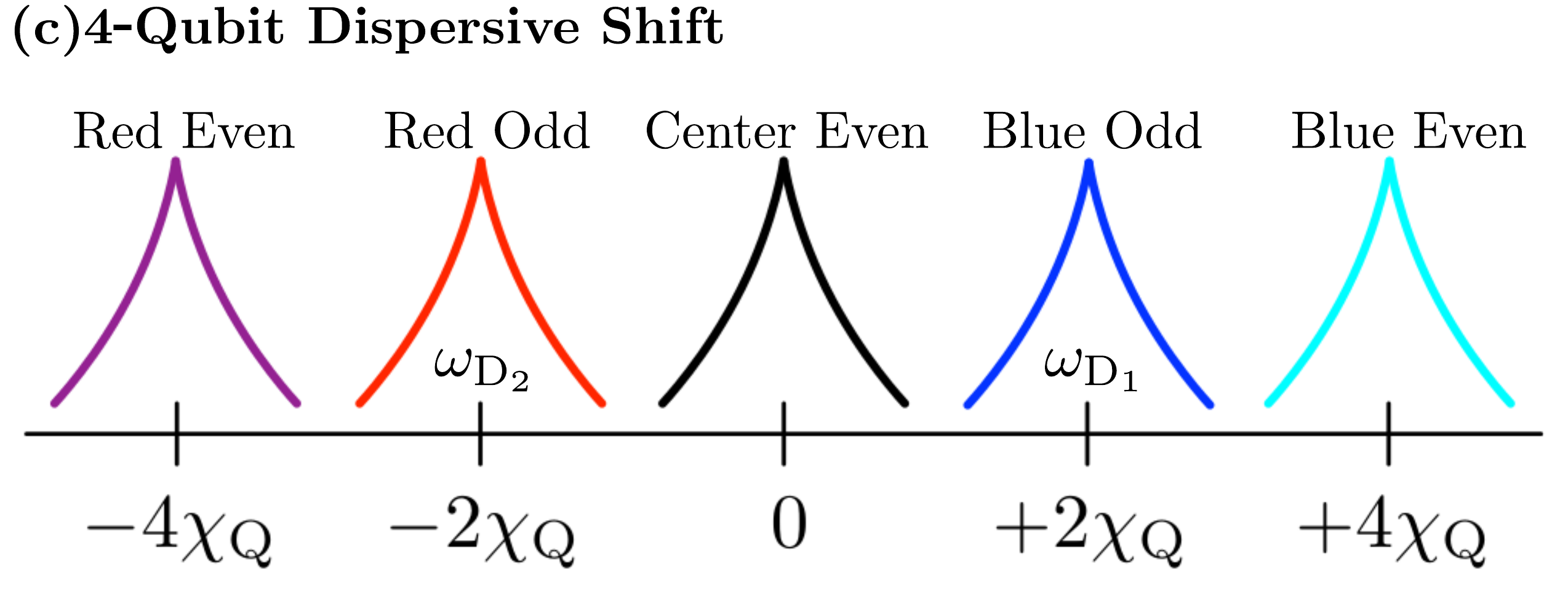}}
\caption{{\bf(a)} Schematic of the experimental design showing the cavity, $N$ coupled qubits, and the photon counter (a JPM in this case). {\bf (b) and \bf (c)} Illustration of the dispersive shifts on the cavity for two- and four-qubit computational basis states.}
\end{figure}

{\it During the drive stage}, the reduced Hamiltonian of the cavity coupled to a single qubit, 
\begin{align}
\hat{H}_{\rm C} =\tilde{\ww}_{\rm C}\hat{a}^{\dagger}\hat{a} + A(t)\left( \hat{a}+ \hat{a}^{\dagger}\right),
\label{eqn:HC}
\end{align}
is that of a single mode oscillator with resonance frequency $\tilde{\ww}_{\rm C}=\omega_{\rm C}+\tilde{\chi}_{\rm Q}$, where $\tilde{\chi}_{\rm Q}= s\chi_{\rm Q}$ encodes the state of the qubit through $s=\pm 1$, with $s$ the respective eigenvalue of $\hat{\sigma}_z$. By applying a carefully timed classical drive at frequency $\omega_{\rm D}= \omega_{\rm C}-{\chi}_{\rm Q}$, we put the cavity in a high-amplitude coherent state if the qubit is in its excited state, but in the vacuum state if the qubit is in its ground state, thus dephasing the qubit.
The cavity is populated conditionally upon the qubit having odd parity (excited state), effectively applying the unitary
\be
\hat{U}_{\rm D} = \hat{P}^{\rm E}_{\rm Q}\otimes\mathbb{I}_{\rm C} + \hat{P}^{\rm O}_{\rm Q}\otimes\hat{D}(\beta),
\label{eqn:drive}
\ee
which transfers qubit parity information to the cavity, where $\hat{D}(\beta)$ is the displacement operator. Here $\hat{P}^{\rm E/O}_{\rm Q}$ are the projectors onto the even/odd parity subspaces. For single-qubit readout these are projectors onto the qubit ground and excited states, while for more than one qubit they are sums of projectors onto a set of states spanning each parity subspace.

Now consider two qubits coupled to the cavity with equal dispersive shifts. For the computational basis states the total dispersive shift on the cavity is $\tilde{\chi}_{\rm Q}=\pm2\chi_{\rm Q}$ when both qubits are in the same eigenstate (even parity), while $\tilde{\chi}_{\rm Q}=0$ when the qubits are in different eigenstates (odd parity), as illustrated in FIG.~\ref{fig:2PBands}. By driving with frequency $\omega_{\rm D}= \omega_{\rm C}$, we entangle the qubit parity subspaces with distinguishable cavity states (as was done for single-qubit readout), implementing the unitary of Eq.~(\ref{eqn:drive}). At the end of the drive pulse $A(t) = a_0\cos{(\ww_{{\rm D}} t+\phi)}\Theta(t_{\rm D}-t)$, the qubit-dependent cavity occupations are
\begin{align}
\nonumber &\abs{\alpha_{\rm E}}^2 = \left(\frac{a_0}{\Delta_{\rm D}}\right)^2\frac{1-\cos{(\Delta_{\rm D}t_{\rm D})}}{2}, 
  \  \abs{\alpha_{\rm O}}^2 = \left(\frac{a_0}{2}t_{\rm D}\right)^2,
\label{eqn:als2}
\end{align}
where $\Delta_{\rm D}=\tilde{\omega}_{\rm C}-\omega_{{\rm D}}$. Setting $\omega_{\rm D}=\omega_{\rm C}$ gives $|\Delta_{\rm D}| = 2\chi_{\rm Q}$ for even-parity states, and therefore $\abs{\alpha_{\rm E}}^2 = 0$ at $t_{\rm D}=\pi/\chi_{\rm Q}$ ($\Delta_{\rm D} = 0$ for the odd states). As $\al_{\rm E} = 0$, the even-parity states are indistinguishable, and thus, while inter-subspace coherence is reduced to $\braket{\al_{\rm O}}{\al_{\rm E}}$, intra-subspace coherence is protected, unlike in direct parity measurement \cite{Riste2013}.

For $N>2$ qubits, the degeneracy within the odd and even subspaces splits and we can no longer perform parity measurement by a single frequency cavity drive. However, with
\be
A(t) = a_0\sum_i\cos{(\ww_{{\rm D}_i} t+\phi)}\quad 0\le t\le t_D
\ee
where $\omega_{{\rm D}_i}=\omega_{\rm C} + \tilde{\chi}_{{\rm Q}_i}$ are the dispersively-shifted cavity frequencies for each band of the odd-parity subspace, we apply the unitary of Eq.~(\ref{eqn:drive}) by simultaneously driving all odd-parity spectral lines resonantly with a multi-tone drive (see the supplementary material for further details).

For four qubits, basis states with odd parity produce dispersive shifts of $\pm2\chi_{\rm Q}$ (blue and red odd-parity bands), while basis states with even parity cause a dispersive shift of $0$ or $\pm4\chi_{\rm Q}$, as shown in FIG.~\ref{fig:PBands}.  Therefore, with the two drive frequencies $\omega_{{\rm D}_{1,2}}=\omega_{\rm C}\pm2\chi_{\rm Q}$ we can simultaneously drive both odd-parity spectral lines. The cavity occupations for the four-qubit parity bands are shown in FIG.~\ref{fig:4Drive} (see the supplement for their analytic form). Crucially, $\abs{\al_{\rm E}}^2 = 0$ for all even-parity states at $t_{\rm D} = \pi/\chi_{\rm Q}$, while $\abs{\al_{\rm O}}^2$ is equal for all odd-parity states. Therefore, qubit parity information has been mapped onto cavity occupation in a way that does not discriminate between states of the same parity, while reducing inter-subspace coherence to $\braket{\al_{\rm O}}{\al_{\rm E}}$.

\begin{figure}
\includegraphics[width = 0.9\columnwidth]{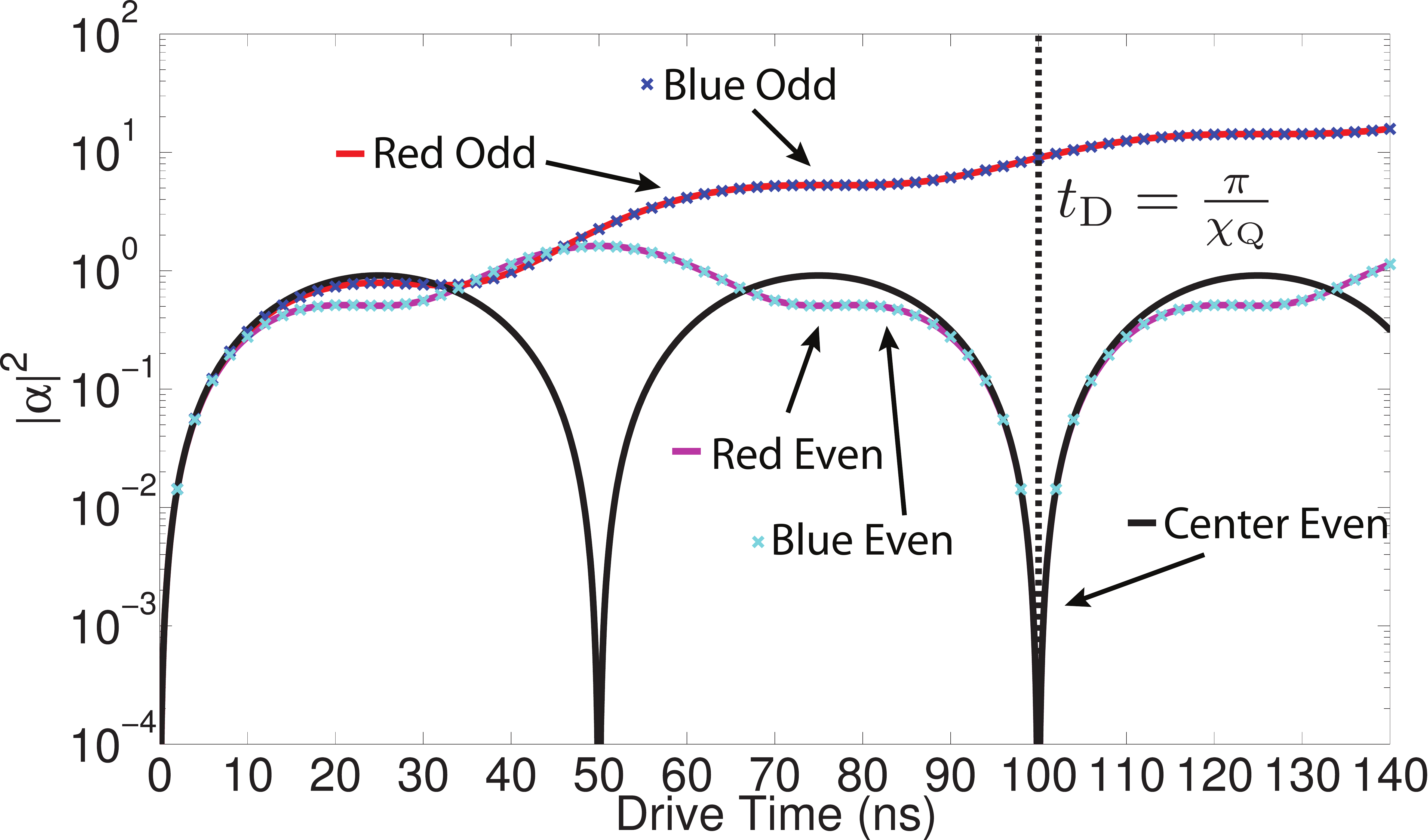}
\caption{Cavity photon number as a function of the length of the drive pulse, $t_{\rm D}$, for the four qubit parity bands. $\chi_{\rm Q}/\pi = 10$ MHz, so the optimal drive time is $t_{\rm D} = 100$ ns.}
\label{fig:4Drive}
\end{figure}

{\it In the measurement stage} we distinguish between the qubit-parity dependent cavity states in a frequency insensitive way (to avoid intra-subspace decoherence) by using a high bandwidth photon counter. We tune the counter on resonance with the bare cavity frequency $\ww_{\rm C}$. For two qubits, the counter and cavity are resonant if the qubits have odd parity, maximizing the detection probability of $\ket{\al_{\rm O}}$. For four qubits, the counter is resonant with the cavity if the qubits are in the center even-parity band; however, this is not an issue because the cavity is unoccupied for even-parity qubits. The counter will be $\pm 2\chi_{\rm Q}$ off resonance from the cavity if the qubits have odd parity, symmetrically between the two odd-parity cavity frequencies. For a counter with bandwidth $> 4\chi_{\rm Q}$ this ensures the detection probabilities for all odd-parity states are nearly identical, minimizing intra-subspace decoherence.

\begin{figure}
\includegraphics[width = \columnwidth]{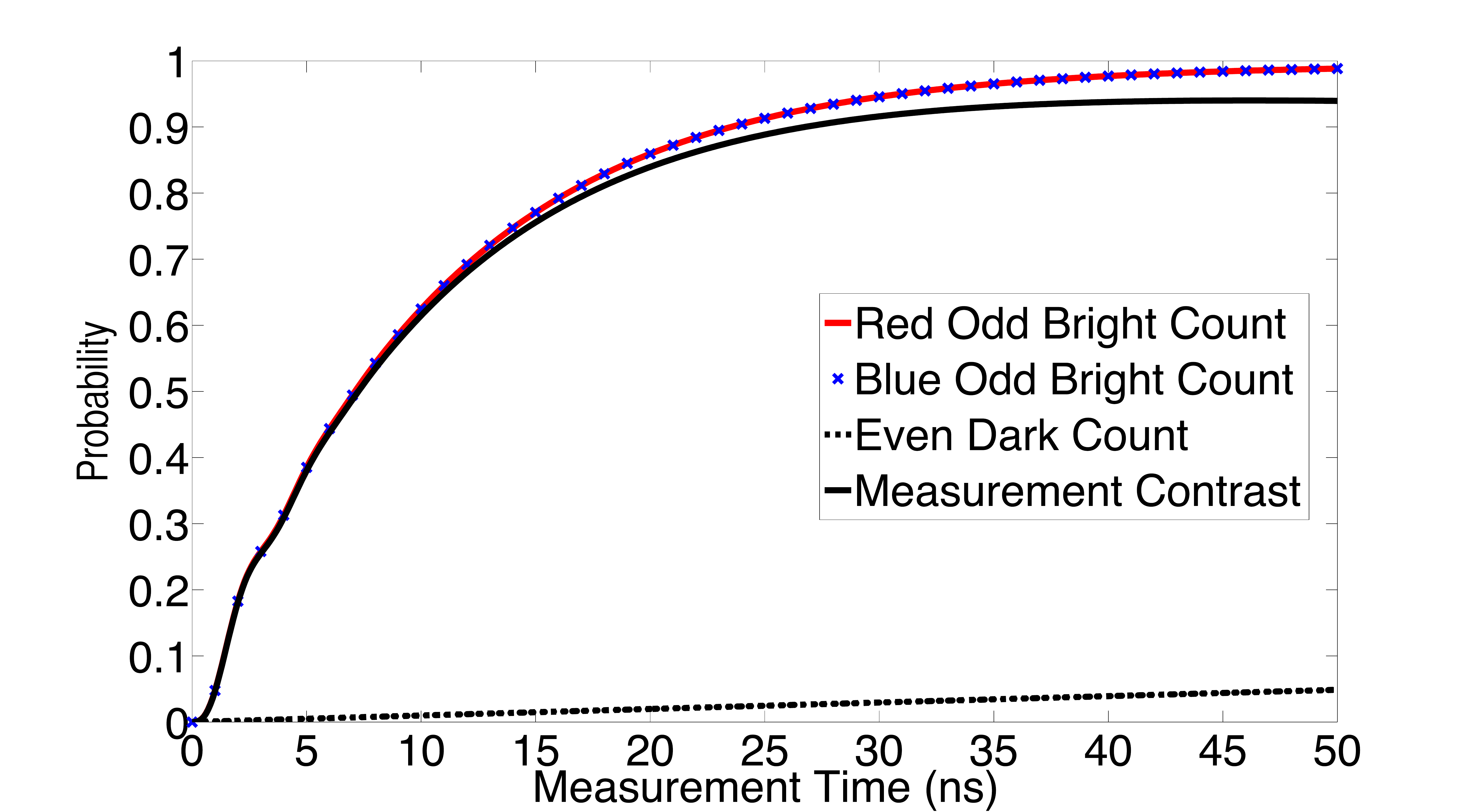}
\caption{Bright count probability for the four-qubit odd-parity bands, dark count probability for even parity, and measurement contrast as functions of measurement time.}
\label{fig:ProbO}
\end{figure}

For $N=4$, the bright count detection probability for the two odd-parity bands is shown in FIG.~\ref{fig:ProbO}, as is the false ``even-state detection probability'' due to dark counts. As expected, the two bright count curves are identical. These curves are calculated by solving the master equation with relevant experimental parameters (for further information see the supplements), where, as an example of a threshold photon counter we have chosen the JPM, which when coupled to a cavity approaches unit detection probability with large bandwidth ($\gg \chi_{\rm Q}$) \cite{Poudel2012}, and is a true dichotomic detector that distinguishes between the vacuum and its complement \cite{Govia2012,Govia2014}. Similar detectors exist in the optical \cite{Dakna:1997uq,Oi:2013fk} and near infrared \cite{Boitier:2009xy} regimes.

As a figure of merit we consider the parity measurement contrast, defined as 
\be
C\left(t_{\rm M}\right) \equiv P\left(\abs{\al_{\rm O}}^2,t_{\rm M} \right) - P\left(\abs{\al_{\rm E}}^2,t_{\rm M} \right),
\ee
the difference in detection probability between odd/even parity states, which is a function of the measurement time $t_{\rm M}$. The drive time $t_{\rm D}$ is chosen such that $\al_{\rm E} =0$, so the parity measurement contrast reduces to the detection probability of the odd-parity cavity state minus the dark count probability. In reality there will be thermal photons in the cavity. However, because of the threshold nature of the photon counter it is not necessary to have an initially empty cavity to maximize measurement contrast; it is sufficient that the detection probability has a sharp threshold between $\al_{\rm O}$ and $\al_{\rm E}$. This is important for the JPM, where, while in principle the threshold can be set at zero occupation \cite{Govia2012,Poudel2012}, this is often difficult to engineer. Nonetheless, a suitable threshold between $\al_{\rm O}$ and $\al_{\rm E}$ is easily obtained \cite{Govia:2014fk}.

The parity measurement contrast is limited by misidentification of the parity, which occurs either due to a dark count or the nonzero probability of not detecting the state $\ket{\al_{\rm O}}$. These errors are controlled by the bright count/dark count ratio of the photon counter and the cavity occupation $\abs{\al_{\rm O}}^2$, and control over these parameters is sufficient to obtain contrast arbitrarily close to unity. As seen in FIG.~\ref{fig:ProbO}, measurement contrast that approaches 95\% is achievable, compatible with the readout threshold for error correction \cite{Fowler:2012sf}\footnote{To the best of our knowledge the imbalanced thresholds between gates and measurement have not been explored in detail in the fault tolerance literature.}, in a 40 ns time-frame with experimentally relevant bright count/dark count ratio and $\abs{\al_{\rm O}}^2$.

{\it In the reset stage} we again implement the qubit-dependent cavity displacement of Eq.~(\ref{eqn:drive}), but with $\beta = -\al_{\rm M}$.  If the qubits have odd parity, the cavity begins approximately in the coherent state $\ket{\al_{\rm M}}$ and is returned to vacuum, where $\abs{\al_{\rm M}}$ can be calculated given the input magnitude $\abs{\al_{\rm O}}$ and the form of the detection back action on the cavity \cite{Govia2014}. For a JPM, after measurement the cavity is not a coherent state, and reset is imperfect. However, single-shot parity measurement is unaffected by reset error, and though repeated measurements can be affected, for experimentally relevant parameters reset error is on the order of 1\% \cite{Govia:2014fk}.

{\it In a realistic experiment} it is unlikely that all qubit-cavity dispersive shifts will be identical. Therefore, we examine the robustness of our protocol under variations in the dispersive shifts, defining the dispersive shift error $\eps$, such that $\chi_{\rm Q_1} = \chi_{\rm Q}$, and $\chi_{\rm Q_2} = \chi_{\rm Q} + \eps$ for two qubits. For four qubits there are three dispersive shift errors: $\eps_2$, $\eps_3$, and $\eps_4$. We assume this error is small, such that $\eps_i/\chi_{\rm Q} \ll 1$. For a superconducting qubit with coherence time 5 $\mu$s, the uncertainty in $\chi_{\rm Q}$ due to homogeneous broadening is no more than $\pm0.1\%$, and thus $\abs{\eps_i/\chi_{\rm Q}} < 0.2\%$. Mismatch in $\chi_{\rm Q}$ causes a reduction in measurement contrast and intra-subspace decoherence (dephasing of a superposition of qubit states of the same parity), and we provide quantitative estimates of these effects, leaving further details to the supplement. Errors in the drive frequencies $\ww_{\rm D_i}$ cause similar errors as $\chi_{\rm Q}$ mismatch, but can be corrected for during tune-up.

\begin{figure}
\includegraphics[width = \columnwidth]{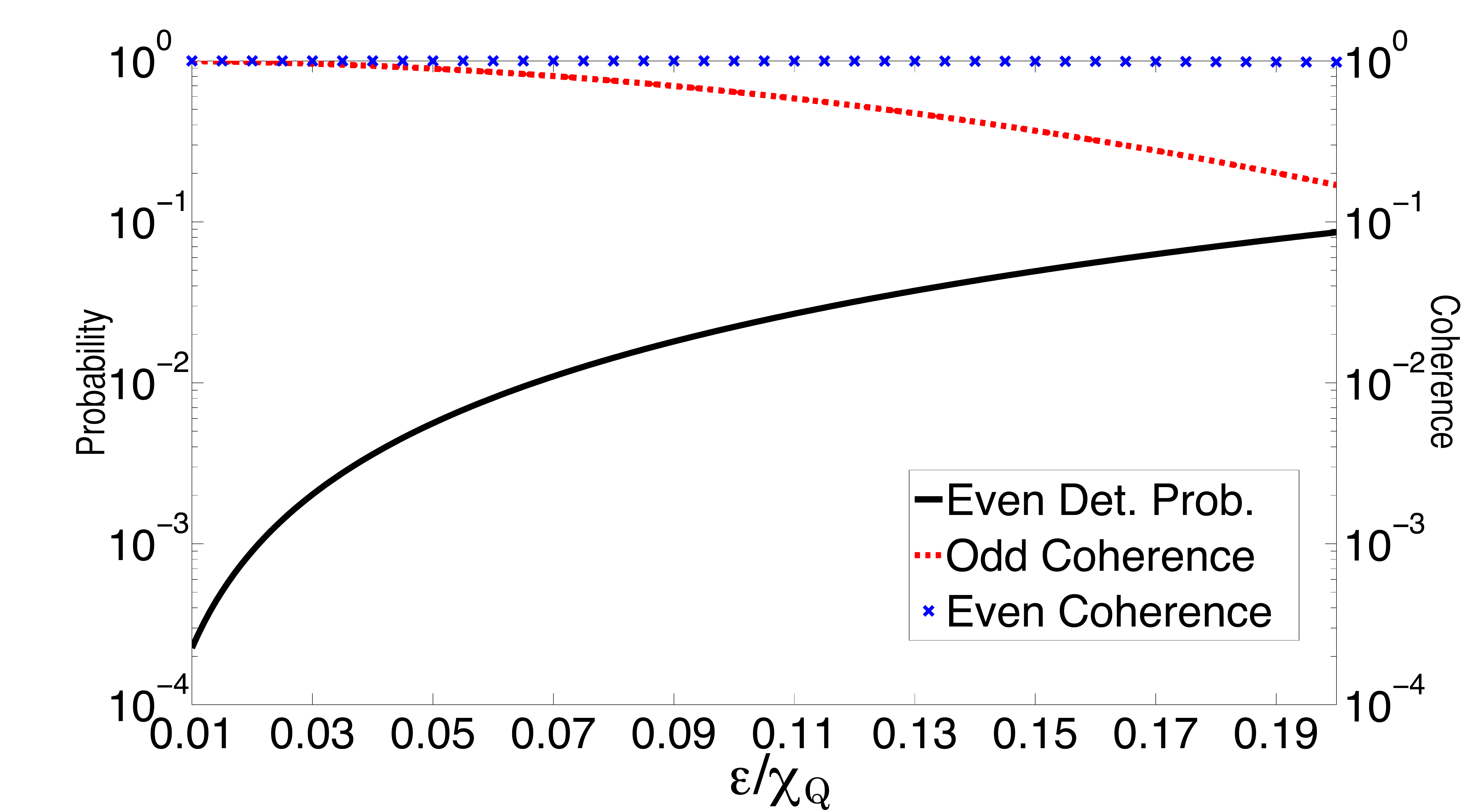}
\caption{Even-parity detection probability, odd-parity coherence, and even-parity coherence as functions of the $\chi_{\rm Q}$ mismatch for two-qubit parity measurement. $\eps/\chi_{\rm Q}$ ranges from 1\% to 20\%, well within experimental expectations. $\chi_{\rm Q}/2\pi = 5$ MHz, and the drive power $a_0$ is such that $\abs{\al_{\rm O}}^2 = 9$.}
\label{fig:ChiMis}
\end{figure}

Imperfect resonance between the applied drives and the cavity due to $\chi_{\rm Q}$ mismatch leads to a reduced $\abs{\al_{\rm O}}^2$ and an increased $\abs{\al_{\rm E}}^2$. For either two or four qubits, the reduction in measurement contrast is second order in $\eps_i/\chi_{\rm Q}$, commensurate with a square-law detector. As shown in FIG.~\ref{fig:ChiMis}, even for 10\% $\chi_{\rm Q}$ mismatch the resulting measurement error (even-state detection probability) is on the order of 1\%.

The intra-subspace decoherence caused by $\chi_{\rm Q}$ mismatch is due to qubit states with the same parity being entangled to cavity states with different phases, making them distinguishable (in principle). The resulting decoherence is at most second order in $\eps_i/\chi_{\rm Q}$. Shown in FIG.~\ref{fig:ChiMis} is the coherence of odd- and even-parity two-qubit superposition states, as quantified by the relevant off-diagonal matrix element of the reduced two-qubit density matrix. Though the decoherence of an odd-parity superposition is nontrivial for larger values of $\eps/\chi_{\rm Q}$, perfect reset returns full coherence by depopulating the cavity in a phase insensitive way.

However, any cavity decay prior to perfect reset will cause irreversible coherence loss, as would the decay of residual photons after imperfect reset. Up to the limit of strong $\chi_{\rm Q}$-mismatch ($\eps$ approaching $\chi_{\rm Q}$), the intra-subspace coherence decays to $\exp(-N_{\rm C}(1-\cos(\pi\eps/\chi_{\rm Q})))$ in the steady state, where $N_{\rm C}$ is the average photon number in the cavity prior to decay, hence requiring $\epsilon<\chi_{\rm Q}/\sqrt{N_{\rm C}}$. For a sufficiently high-$Q$ cavity, decay during the short measurement time is unlikely, and post-reset cavity decay is a more significant source of loss. For the parameters considered here, the residual photons after imperfect reset result in coherence loss that is less than $1\%$ in the worst case, and is therefore inconsequential.

For two qubits with $\chi_{\rm Q}$ mismatch, no intra-subspace decoherence occurs during the measurement stage as cavity states corresponding to qubit states of the same parity have equal detection probability. For four qubits with $\chi_{\rm Q}$ mismatch the detection probabilities for states within the same parity subspace can be different, with their difference scaling as $(\eps_i^2-\eps_j^2)/\chi_{\rm Q}^2$. This changes the basis of measurement, which changes the output qubit state at the end of the protocol. This effect can be mitigated by a square-law photon counter that quickly saturates with photon number above threshold, such as the JPM, for which the error in the output qubit state is negligible.

Higher order effects beyond the dispersive Hamiltonian can affect the parity readout protocol presented here. As a result of the formation of cavity-qubit dressed states, there will be residual cavity occupation for even qubit states, which reduces parity measurement contrast. For four qubits the contrast is reduced to $\approx 92\%$  (see supplement). However, contrast can be improved by increasing cavity-qubit detuning, and/or using better drive pulse sequences. The protocol's QNDness will also be affected by higher order terms; however, as shown for single-qubit readout in \cite{Govia:2014fk}, this change in QNDness is minimal. As the measurement stages of single-qubit readout and parity readout are effectively identical, this is also true for parity readout.

Our protocol involves an entangling operation that maps qubit information to the cavity such that destructive readout of the cavity non-destructively determines the qubits' parity. It provides a tailored and efficient quantum circuit that maps parity information onto an ancilla (the cavity), and then measures the ancilla in a way that is insensitive to qubit-resolving information. It occupies a middle ground between the direct, amplifier-based parity measurement of \cite{DiVincenzo2013,Tornberg:2014nr} and gate-based parity measurement. The advantage of our protocol over gate-based protocols is the reduced number of entangling gates required: one for our proposal {\it versus} four for the traditional gate-based four-qubit protocol. Our protocol also requires fewer cavities than direct four-qubit parity readout \cite{DiVincenzo2013,Tornberg:2014nr}.

{\it In conclusion} we have presented a high-fidelity, scalable, QND protocol for parity readout of two or four qubits via photon counting of a dispersively coupled cavity. Measurement contrast is limited by the bright/dark count ratio of the photon counter, and with the limited optimization in this Letter approaches 95\%. Our protocol introduces no decoherence of qubit states with the same parity, and is robust against the major sources or error. Future work will focus on protocol optimization and higher order corrections.

\begin{acknowledgments}
Supported by the Army Research Office under contract W911NF-14-1-0080. L.C.G.G., E.J.P., and F.K.W. also acknowledge support from the European Union through ScaleQIT and LCGG from NSERC through an NSERC PGS-D. M.G.V., and R.M. were also supported by NSF Grant No. DMR-1105178.
\end{acknowledgments}

\bibliography{Parity}
\cleardoublepage


\section{Supplementary Material}
In the following supplemental sections we expand on the details of the calculations and simulations presented in the main text. While not essential  to understanding the results presented in the main text, these details complete the explanation of the results in a more mathematically rigorous way.

In section \ref{sec:Sd4} we present the details of the analytic derivation of FIG. 2 of the main text, which shows the driven cavity occupation as a function of time, and how this is used to create a qubit-parity dependent cavity population. In section \ref{sec:SME} we describe the master equation formalism used to simulate the interaction of the cavity with our example threshold square-law photon counter, the Josephson Photomultiplier (JPM). In section \ref{sec:SChi} we expand on the error analysis of the main text to describe the effect of $\chi_{\rm Q}$ mismatch. In section \ref{sec:decay} we discuss the effect of this $\chi_{\rm Q}$ mismatch when cavity photon loss occurs. Finally, in section \ref{sec:SHigh} we describe how consideration of the full Jaynes-Cummings Hamiltonian affects the results of the main text.

\section{Derivation of the qubit-parity dependent drive for four qubits}
\label{sec:Sd4}

In this section we derive the driven cavity evolution for four qubits, as shown in FIG. 2 of the main text. The full system Hamiltonian is block diagonal, with one block for each state of the qubits. In each block the effective cavity Hamiltonian is 
\begin{align}
\nonumber\hat{H}_{\rm C} = &\tilde{\ww}_{\rm C}\hat{a}^{\dagger}\hat{a} + a_0[\cos{(\ww_{\rm D_1} t+\phi_1)} \\
&+\cos{(\ww_{\rm D_2} t+\phi_2)}]\Theta(t_{\rm D}-t)\left( \hat{a}+ \hat{a}^{\dagger}\right),
\label{eqn:apHC}
\end{align}
where the shifted cavity frequency $\tilde{\ww}_{\rm C}=\omega_{\rm C}+\tilde{\chi}_{\rm Q}$ depends on the qubit-state dependent dispersive shift $\tilde{\chi}_{\rm Q}=\chi_{\rm Q}\big<\hat{\sigma}_z\big>$, which defines the state of the qubits, and therefore, the corresponding block of the full Hamiltonian. $\tilde{\chi}_{\rm Q} = \pm 2 \chi_{\rm Q}$ for odd-parity states and $\tilde{\chi}_{\rm Q} = 0, \pm 4 \chi_{\rm Q}$ for even parity.

We go to a frame rotating with the shifted cavity frequency $\tilde{\ww}_{\rm C}$ to obtain
\begin{align}
\hat{H}_{\rm C}' = \frac{a_0}{2}\Theta(t_{\rm D}-t)\left[ \hat{a}\left( e^{-i\Delta_1t} +  e^{-i\Delta_2t}\right) + {\rm h.c.}\right],
\label{eqn:Hrot}
\end{align}
where $\Delta_i = \tilde{\ww}_{\rm C} - \ww_{\rm D_i}$. We formally solve for the evolution operator in each block for times $t \geq t_D$ as
\be
\hat{U}(t,0) = \mathcal{T} {\rm exp}\left\{ -i\int_0^{t_{\rm D}} \hat{H}_{\rm C}'(t') dt'\right\},
\label{eqn:FS}
\ee
where $\mathcal{T}$ is the time ordering operator. The evolution operator for the full system then has the form
\be
\hat{U}_{\rm D} = \sum_j \ket{j}\bra{j}\otimes\hat{U}_{j}(t,0),
\ee
where $\hat{U}_{j}(t,0)$ are the solutions to equation (\ref{eqn:FS}) for each qubit state $\ket{j}$ in the computational basis.

We evaluate the integral in equation (\ref{eqn:FS}) using the Magnus expansion \cite{Magnus:1954yu} to take care of the time ordering. Because $\left[\hat{a},\hat{a}^\dagger\right] = 1$ the Magnus expansion truncates at second order, and conveniently the second order term is a global phase that we can ignore. We evaluate the integral in two regimes, corresponding to even or odd parity, and in so doing obtain the solution for the blocks of the full Hamiltonian. For even parity, $\Delta_1 \neq 0$ and $\Delta_2 \neq 0$, and to first order we have
\begin{align}
\nonumber \int_0^{t_{\rm D}} \hat{H}_{\rm C}'(t') dt' =& \frac{a_0}{2}\Big[ \hat{a}\Big( \frac{i}{\Delta_1}\left( e^{-i\Delta_1{t_{\rm D}}} - 1 \right)\\
\nonumber &+ \frac{i}{\Delta_2}\left( e^{-i\Delta_2{t_{\rm D}}} -1 \right)\Big) + {\rm h.c.}\Big].
\end{align}
The evolution operator of equation (\ref{eqn:FS}) for even parity is then
\be
\hat{U}_{\rm E}(t,0) = \hat{D}(\al_{\rm E}(t_{\rm D})),
\label{eqn:unitary}
\ee
where
\begin{align}
\al_{\rm E}(t_{\rm D}) = &-\frac{a_0}{2}\left[ \frac{e^{i\Delta_1{t_{\rm D}}} - 1}{\Delta_1}+ \frac{e^{i\Delta_2{t_{\rm D}}} -1 }{\Delta_2}\right].
\end{align}
In general $\al_{\rm E}(t_{\rm D})$ varies in phase for different even-parity qubit states as $\Delta_{1,2}$ depend on the specific state.

For an odd-parity qubit basis state, the shifted cavity will be on resonant with one of the two applied drives, and therefore, one $\Delta_i$ is zero, while the other is nonzero. Since the solution is symmetric in regards to which $\Delta_i = 0$, we will examine $\Delta_1 = 0$ and $\Delta_2 \neq 0$ without loss of generality. The integral in equation (\ref{eqn:FS}) now gives
\begin{align}
\nonumber\int_0^{t_{\rm D}} \hat{H}_{\rm C}'(t') dt' = \frac{a_0}{2}\left[ \hat{a}\left( t_{\rm D} + i\frac{e^{-i\Delta_2{t_{\rm D}}} -1 }{\Delta_2}\right) + {\rm h.c.}\right],
\end{align}
which leads to the odd-parity evolution operator
\be
\hat{U}_{\rm O}(t,0) = \hat{D}(\al_{\rm O}(t_{\rm D})) = \hat{D}\left(\beta(t_{\rm D}) + \eta(t_{\rm D})\right),
\ee
where
\begin{align}
&\beta(t_{\rm D}) = -\frac{ia_0}{2}t_{\rm D} \\
&\eta(t_{\rm D}) = -\frac{a_0}{2\Delta_2}\left(e^{i\Delta_2{t_{\rm D}}} -1\right).
\end{align}
In this case it is important to note that while $\eta(t_{\rm D})$ varies between odd-parity states, $\beta(t_{\rm D})$ is the same for all odd-parity states.

If $\ww_{\rm D_1} = 2\chi_{\rm Q}$ and $\ww_{\rm D_2} = -2\chi_{\rm Q}$ as in the main text, then for even parity we have $\Delta_1,\Delta_2 \in \left\{\pm2\chi_{\rm Q},\pm6\chi_{\rm Q} \right\}$. For odd parity, we have chosen $\Delta_1 = 0$, which means that $\Delta_2 = 4\chi_{\rm Q}$. Thus, if we set $t_{\rm D} = \pi/\chi_{\rm Q}$, we have that $\al_{\rm E}(t_{\rm D}) = 0$, and $\eta(t_{\rm D}) = 0$, while $\beta(t_{\rm D})  \neq 0$. As a result of this,
\begin{align}
&\hat{U}_{\rm E}(t,0) = \hat{D}(0) = \mathbb{I}, \\
&\hat{U}_{\rm O}(t,0) = \hat{D}\left(\beta(t_{\rm D})\right),
\end{align} 
and the evolution operators are the same for all states within the same parity subspace. This, along with the block diagonal form of the full system Hamiltonian implies that the unitary on the full system will have the form
\begin{align}
\nonumber\hat{U}_{\rm D} &=  \sum_{j\in{\rm even}} \ket{j}\bra{j}\otimes\hat{U}_{\rm E}(t,0) +  \sum_{j\in{\rm odd}} \ket{j}\bra{j}\otimes\hat{U}_{\rm O}(t,0) \\
&= \hat{P}^{\rm E}_{\rm Q}\otimes \mathbb{I}_{\rm C} +\hat{P}^{\rm O}_{\rm Q}\otimes \hat{D}(\beta),
\label{eqn:drive}
\end{align}
where $ \hat{P}^{\rm E,O}_{\rm Q}$ are projectors onto the even- and odd-parity subspaces respectively.

\section{Master equation for the cavity-JPM coupled system}
\label{sec:SME}

In this section we describe the methodology of the numerical simulations used to obtain FIG. 3 of the main text, which shows the detection probability and parity contrast for our example threshold detector, the JPM.

We start with the cavity-JPM coupled system for four qubit odd parity, where the shifted cavity frequency $\tilde{\ww}_{\rm C} = \ww_{\rm C} \pm 2\chi_{\rm Q}$ and the JPM frequency $\ww_{\rm J} = \ww_{\rm C}$ are such that the JPM is $\pm 2\chi_{\rm Q}$ detuned from the cavity. This leads to the Hamiltonian
\begin{align}
\nonumber\hat{H}_{\rm JC} = \tilde{\ww}_{\rm C}\hat{a}^{\dagger}\hat{a} - \frac{\ww_{\rm J}}{2}\hat{\sigma}_{\rm J} + g_{\rm J}\left(\hat{a}\hat{\sigma}_{\rm J}^{+} + \hat{a}^{\dagger}\hat{\sigma}_{\rm J}^{-}\right),
\label{eqn:HamJC}
\end{align}
where $\hat{\sigma}_{\rm J}^{\pm}$ couple the ground and excited state
of the JPM.  As shown previously \cite{Chen2011,Peropadre:2011fj,Poudel2012,Govia2012}, the JPM is well-approximated as a three-level system with self-Hamiltonian $H_{\rm J}=-\frac{\omega_{\rm J}}{2}\sigma_{\rm J}$, where $\hat{\sigma}_{\rm J} \equiv {\rm diag}(1,-1,k)$. The energy of the third `measurement' state is arbitrary in our model; the JPM only tunnels into it incoherently at fixed rates \cite{Govia2012,Poudel2012}.

In addition, the JPM experiences a number of incoherent processes. Tunneling from the JPM excited state to the measured state corresponds to photon detection and occurs at the bright count rate $\yy_{\rm J}$, with the corresponding Lindblad operator
\be
\hat{L}_{\rm 2} = \sqrt{\yy_{\rm J}}\left( \mathbb{I}_{\rm C} \otimes \ket{\rm m}\bra{1}_{\rm J}\right).
\ee
Inelastic relaxation takes the JPM from the excited state to the ground state at a rate $\yy_{\rm R}$, with corresponding Lindblad operator
\be
\hat{L}_{\rm 1} = \sqrt{\yy_{\rm R}}\left( \mathbb{I}_{\rm C} \otimes \ket{0}\bra{1}_{\rm J}\right).
\ee
Finally, false detections, where the JPM tunnels from the ground state to the measured state, can occur at the dark count rate $\yy_{\rm D}$, with Lindblad operator
\be
\hat{L}_{\rm 0} = \sqrt{\yy_{\rm D}}\left( \mathbb{I}_{\rm C} \otimes \ket{m}\bra{0}_{\rm J}\right).
\ee
We ignore the effects of pure dephasing on the JPM as they do not affect the parity measurement protocol of the main text.

We solve the master equation
\begin{align}
\nonumber \dot{\rho}(t) = &-i[\hat{H}_{\rm JC},\rho(t)] \\
&+ \sum_{\mu=0}^2\left(\hat{L}_\mu\rho(t)\hat{L}_\mu^\dagger-\frac{1}{2}\{ \hat{L}_\mu^\dagger \hat{L}_\mu,\rho(t)\}\right),
\label{eqn:ME}
\end{align}
with $\yy_{\rm J} = 200$ MHz, $\yy_{\rm R} = 200$ MHz, $\yy_{\rm D} = 1$ MHz, and $g_{\rm J}/2\pi = 50$ MHz, for both $\tilde{\ww}_{\rm C} - \ww_{\rm C}  = 2\chi_{\rm Q}$, and $-2\chi_{\rm Q}$ (where $\chi_{\rm Q}/2\pi = 5$ MHz) to generate the detection probability curves of the main text FIG. 3.

\section{Qubit Dispersive Shift Mismatch}
\label{sec:SChi}

In this section we examine in detail the possible sources of error caused by $\chi_{\rm Q}$ mismatch that were described in the main text and shown in FIG. 4. Firstly, we quantify the decrease in measurement contrast and the coherence loss for two qubits. As described in the main text, this coherence loss can be restored by perfect reset. We then examine these effects for four qubits, and also quantify the change in the measurement basis due to detection probability mismatch, which is unique to four qubits.

\subsection{Two Qubits}
\label{sec:2Q}

If the dispersive shifts for the two qubits vary, such that $\chi_{\rm Q_1} =\chi_{\rm Q}$ and $\chi_{\rm Q_2} =\chi_{\rm Q} + \eps$, then the cavity-drive detunings within a parity subspace split, such that we now have
\begin{align}
\nonumber&\Delta_{00} = -2\chi_{\rm Q} - \eps \\
\nonumber&\Delta_{11} = 2\chi_{\rm Q} + \eps \\
\nonumber&\Delta_{01} =  \eps \\
&\Delta_{10} =  - \eps
\end{align}  
where $\Delta_{ij}$ is the cavity-drive detuning for the state $\ket{ij}$.

After a time $t_{\rm D} = \pi/\chi_{\rm Q}$ the cavity will be in a qubit-state dependent coherent state with amplitude
\begin{align}
\nonumber&\al_{00} = \frac{a_0}{2(2\chi_{\rm Q}+\eps)}\left( e^{-i\frac{\pi}{\chi_{\rm Q}}\eps} - 1\right), \\
\nonumber&\al_{11} = -\frac{a_0}{2(2\chi_{\rm Q}+\eps)}\left( e^{i\frac{\pi}{\chi_{\rm Q}}\eps} - 1\right), \\
\nonumber&\al_{01} = -\frac{a_0}{2\eps}\left( e^{i\frac{\pi}{\chi_{\rm Q}}\eps} - 1\right), \\
&\al_{10} = \frac{a_0}{2\eps}\left( e^{-i\frac{\pi}{\chi_{\rm Q}}\eps} - 1\right).
\label{eqn:als}
\end{align}
For the even-parity qubit states we have that $\al_{11} = - \bar{\al}_{00}$, and for the odd-parity states $\al_{10} = - \bar{\al}_{01}$. As a result, states within the same parity subspace will have the same detection probability, and so there will be no intra-subspace decoherence during the measurement stage. However, as $\abs{\al_{00}} \neq 0$ there will be a reduction of measurement contrast due to the increased probability of misidentification. As the coherent states for each parity subspace are out of phase with one another there also will be intra-subspace decoherence during the drive stage. We will now quantify these effects in the $\eps/\chi_{\rm Q} <1$ regime.

For the measurement contrast and misidentification we can quantify the effect of $\chi_{\rm Q}$ mismatch by looking at the cavity occupation for odd parity, $\abs{\al_{\rm O}}^2=\abs{\al_{01}}^2=\abs{\al_{10}}^2$, and for even parity, $\abs{\al_{\rm E}}^2=\abs{\al_{00}}^2=\abs{\al_{11}}^2$. For odd parity
\begin{align}
\nonumber &\abs{\al_{\rm O}}^2 = \left(\frac{a_0}{2\eps}\right)^2\left(2 - 2\cos{\left(\frac{\pi \eps}{\chi_{\rm Q}}\right)} \right) \\
&= \left(\frac{a_0}{2}\right)^2\left(\frac{\pi}{\chi_{\rm Q}}\right)^2\left(1-\frac{1}{12}\left(\frac{\pi\eps}{\chi_{\rm Q}}\right)^2\right) + \mathcal{O}\left(\varepsilon^4\right),
\end{align}
where $\varepsilon=\eps/\chi_{\rm Q}$, and we see that the reduction of $\abs{\al_{\rm O}}^2$ from that when the dispersive shifts are perfectly matched is second order in $\eps/\chi_{\rm Q}$. Similarly, for even parity
\begin{align}
\nonumber &\abs{\al_{\rm E}}^2 = \left(\frac{a_0}{2(2\chi_{\rm Q} +\eps)}\right)^2\left(2 - 2\cos{\left(\frac{\pi\eps}{\chi_{\rm Q}}\right)} \right) \\
\nonumber &= \left(\frac{a_0}{2(2\chi_{\rm Q} +\eps)}\right)^2\left(\frac{\pi\eps}{\chi_{\rm Q}}\right)^2 + \mathcal{O}\left(\varepsilon^4\right)\\
&= \left(\frac{a_0}{2}\right)^2\left(\frac{\pi}{\chi_{\rm Q}}\right)^2\left(\frac{\eps}{2\chi_{\rm Q}}\right)^2 + \mathcal{O}\left(\varepsilon^4\right),
\end{align}
and the increase in $\abs{\al_{\rm E}}^2$ from that when the dispersive shifts are perfectly matched is again second order in $\eps/\chi_{\rm Q}$. The fact that the lowest order dependence is quadratic originates from the expansion of the cosine, and can be physically explained by the fact that the amplitude is originally tuned to zero, and a square-law detector responds to intensity which is $\abs{\al}^2$.

To quantify the intra-subspace decoherence during the drive stage, we consider an arbitrary superposition of odd-parity states with the cavity initially in vacuum, to which we apply the modified qubit state dependent drive (accounting for $\chi_{\rm Q}$ mismatch). The resulting state 
\be
\ket{\Psi} = a\ket{01}\ket{\al_{01}} + b\ket{10}\ket{\al_{10}},
\ee
where $\abs{a}^2+\abs{b}^2=1$, is no longer a product state and as such there will necessarily be decoherence of the reduced qubit state. The reduced qubit state is
\begin{align}
\rho_{\rm Q} = \left( \begin{array}{cccc}
0 & 0 & 0 & 0 \\
0 & \abs{a}^2 & \bar{D}a\bar{b} & 0 \\
0 & D\bar{a}b & \abs{b}^2 & 0 \\
0 & 0 & 0 & 0
\end{array}\right),
\end{align}
with the complex decoherence factor
\be
D = \braket{\al_{01}}{\al_{10}} = \exp{\left\{-\left(\abs{\al_{01}}^2 + \bar{\al}_{01}^2 \right)\right\}}.
\ee
From equation (\ref{eqn:als}) we see that
\begin{align}
&\abs{\al_{01}}^2 = 2A_{\rm O}^2\left(1 - \cos{\left(\frac{\pi\eps}{\chi_{\rm Q}}\right)} \right), \\
&\nonumber\bar{\al}_{01}^2 = A_{\rm O}^2\bigg[1 - 2\left(\cos{\left(\frac{\pi\eps}{\chi_{\rm Q}}\right)} -i\sin{\left(\frac{\pi\eps}{\chi_{\rm Q}}\right)}\right) \\
&+\cos{\left(2\frac{\pi\eps}{\chi_{\rm Q}}\right)} -i\sin{\left(2\frac{\pi\eps}{\chi_{\rm Q}}\right)} \bigg],
\end{align}
where $A_{\rm O} = a_0/2\eps$.

Keeping only the first nonzero term in the Taylor series expansion we
find
\begin{align}
&\abs{\al_{01}}^2 + \bar{\al}_{01}^2 = A_{\rm O}^2\left(\frac{1}{2}\left(\frac{\pi\eps}{\chi_{\rm Q}}\right)^4 +i\left(\frac{\pi\eps}{\chi_{\rm Q}}\right)^3 \right) \label{eqn:OD} \\
\nonumber&= \left(\frac{a_0}{2}\right)^2\left(\frac{\pi}{\chi_{\rm Q}}\right)^2\left(\frac{1}{2}\left(\frac{\pi\eps}{\chi_{\rm Q}}\right)^2 +i\frac{\pi\eps}{\chi_{\rm Q}}\right) + \mathcal{O}\left(\varepsilon^3\right).
\end{align}
As can be seen, for an odd-parity superposition, $\chi_{\rm Q}$ mismatch causes decoherence at a rate that is second order in the small parameter $\eps/\chi_{\rm }$, and also introduces a complex phase factor that is first order in $\eps/\chi_{\rm }$.

For an even-parity superposition, we simply replace $A_{\rm O}$ with the corresponding expression for even states, and we find
\begin{flalign}
\nonumber&\abs{\al_{00}}^2 + \bar{\al}_{00}^2 =\left(\frac{a_0}{2(2\chi_{\rm Q} + \eps)}\right)^2\left(\frac{1}{2}\left(\frac{\pi\eps}{\chi_{\rm Q}}\right)^4 +i\left(\frac{\pi\eps}{\chi_{\rm Q}}\right)^3\right) \\
&= \left(\frac{a_0}{4\chi_{\rm Q}}\right)^2\left(\frac{1}{2}\left(\frac{\pi\eps}{\chi_{\rm Q}}\right)^4 +i\left(\frac{\pi\eps}{\chi_{\rm Q}}\right)^3\right) + \mathcal{O}\left(\varepsilon^5\right).
\label{eqn:ED}
\end{flalign}
Thus the even parity situation is even better, as the decoherence rate is now fourth order in $\eps/\chi_{\rm }$, and the complex phase factor now third order.

For both odd- and even-parity superpositions full intra-subspace coherence will be returned to the two qubits by perfect reset, as it disentangles the qubits and the cavity in a unitary and phase insensitive way. However, if cavity decay occurs before reset, or imperfect reset leaves residual photons which subsequently decay, the qubit state will lose coherence, as will be discussed in section \ref{sec:decay}.

\subsection{Four Qubits}

The situation is considerably more complex in the four qubit case as the cavity-drive detunings split into sixteen distinct frequencies, one for each qubit state. Correspondingly, after the qubit state dependent drive there are sixteen possible cavity states $\ket{\al_{ijkl}}$, where $i,j,k,l \in \{0,1\}$ index the qubit state $\ket{ijkl}$. As a result of the double frequency drive each $\al_{ijkl}$ will have two components oscillating at different frequencies, with each component similar in form to those of equation (\ref{eqn:als}). Cavity states for even qubit parity will have both components of the form of $\al_{00}$, where the amplitude is suppressed by a factor $1/(2\chi_{\rm Q}+\eps)$, as both drives are off resonance by at least $2\chi_{\rm Q}$. For odd qubit parity the cavity state will have one component that is similar in form to $\al_{00}$ from the off resonance drive and one component similar to $\al_{01}$ from the nearly on resonance drive.

Given the similar structure of $\al_{ijkl}$ in the four qubit case to $\al_{ij}$ of the two qubit case it is clear that the errors introduced by $\chi_{\rm Q}$ mismatch in the four qubit case will be a generalization of those in the two qubit case. There will be a reduction of measurement contrast due to increase in the probability of misidentification, which will again be at most second order in the small parameters $\eps_i/\chi_{\rm Q}$. The overlap between any two states $\ket{\al_{ijkl}}$ corresponding to qubits in the same parity subspace is very similar to equations (\ref{eqn:OD}) and (\ref{eqn:ED}) for the two qubit case, and as a result, the intra-subspace decoherence during the drive phase will be at most second order in $\eps_i/\chi_{\rm Q}$ for odd-parity superpositions and fourth order for even parity. As in the two qubit case full coherence will be returned to the qubits by perfect cavity reset.

Despite the $\chi_{\rm Q}$ mismatch the two qubit case exhibits symmetry in $\al_{ij}$ within a parity subspace that ensures the detection probabilities are the same for cavity states corresponding to qubit states of the same parity. However, in the four qubit case this symmetry is no longer present, as in general $\eps_2 \neq \eps_3 \neq \eps_4$, and as a result, detection probabilities will differ within a parity subspace.

For example, consider an equal superposition of odd-parity states $\ket{a}$ and $\ket{b}$, which after the drive stage are entangled to cavity stages $\ket{\al_{a/b}}$ respectively. Detection by the photon counter results in back action on the cavity described by the non-unitary back action operator $\hat{B}$. Defining the normalized states
\be
\ket{\psi_{a,b}} = \frac{\hat{B}\ket{\al_{a,b}}}{\bra{\al_{a,b}}\hat{B}^\dagger\hat{B}\ket{\al_{a,b}}} = \frac{\hat{B}\ket{\al_{a,b}}}{P_{a,b}},
\ee
where $P_{a,b}$ is the detection probability of the state $\ket{\al_{a,b}}$, then the resulting state of the cavity-qubits system after detection by the photon counter is
\be
\ket{\Psi} = \frac{1}{\sqrt{N}}\left(\sqrt{P_a}\ket{a}\ket{\psi_a} + \sqrt{P_b}\ket{b}\ket{\psi_b} \right),
\label{eqn:detprobmis}
\ee
where the normalization factor $N = P_a + P_b$.

The four qubit states of equation (\ref{eqn:detprobmis}) are entangled to distinguishable cavity states such that the reduced four qubit density matrix's coherence is reduced by the overlap $\braket{\psi_b}{\psi_a}$. To remove this effect and focus solely on detection probability mismatch we assume that we can perform perfect reset of the cavity and create the state
\be
\ket{\Psi} = \frac{1}{\sqrt{N}}\left(\sqrt{P_a}\ket{a} + \sqrt{P_b}\ket{b} \right)\ket{0},
\ee
for which the reduced qubit state is the pure state $\ket{\Psi}_{\rm Q} = \left(\sqrt{P_a}\ket{a} + \sqrt{P_b}\ket{b} \right)/\sqrt{N}$. Clearly, the output state  $\ket{\Psi}_{\rm Q}$ is no longer the input state and this change is a result of the fact that different detection probabilities for states within the same parity subspace change the basis of measurement of the protocol. To quantify this effect we calculate the magnitude of the overlap between the target state $\ket{\Psi_{\rm T}}_{\rm Q} = \left(\ket{a} + \ket{b} \right)/\sqrt{2}$ and the state $\ket{\Psi}_{\rm Q}$, given by
\be
O = \abs{\braket{\Psi_{T}}{\Psi}_{\rm Q}}^2 = \frac{1}{2}\left(1 + \frac{2\sqrt{P_a}\sqrt{P_b}}{P_a + P_b} \right),
\ee
which is unity for $P_a = P_b$ as expected.

If we assume the photon counter is a JPM with a subtraction operator back action \cite{Govia2014}, then the detection probability of the state $\ket{\al_i}$ is 
\be
P_{i} = 1 - \exp\left\{-\abs{\al_i}^2\right\}.
\ee
By Taylor expanding the {overlap} and discarding terms smaller than $\exp\left(-2\abs{\al_{a/b}}^2\right)$ we obtain the approximate overlap
\begin{flalign}
\nonumber O = &1 - \frac{e^{-\left(\abs{\al_a}^2 + \abs{\al_b}^2\right)}\left({\rm cosh}\left(\abs{\al_a}^2 - \abs{\al_b}^2\right)-1\right)}{4\left(2-e^{-\abs{\al_a}^2} - e^{-\abs{\al_b}^2}\right)} \\
&+ \mathcal{O}\left(e^{-3\abs{\al_{a/b}}^2}\right).
\end{flalign}

Since we are interested in the effect of detection probability mismatch we can assume $\al_{a/b}$ differ in magnitude only and set $\abs{\al_a} = \abs{\al}$ and $\abs{\al_b} = \abs{\al}(1+\delta)$ without loss of generality. The overlap can then be rewritten as
\be
O = 1 - \frac{e^{-\abs{\al}^2\left(2+2\delta +\delta^2\right)}\left({\rm cosh}\left(\abs{\al}^2(2\delta+\delta^2)\right)-1\right)}{4\left(2-e^{-\abs{\al}^2} - e^{-\abs{\al}^2(1+\delta)^2}\right)}.
\ee
To examine the scaling of the overlap, we assume that $\delta \ll 1$ and only keep terms up to order $\delta^2$ in the Taylor series, which gives
\be
O = 1 - \frac{\delta^2\abs{\al}^4e^{-2\abs{\al}^2}}{4\left(1- e^{-\abs{\al}^2}\right)} + \mathcal{O}\left(\delta^3\right).
\label{eqn:detprobO}
\ee
Four qubit $\chi_{\rm Q}$ mismatch causes a difference in the cavity coherent state amplitude for states in the same parity subspace that scales as $\delta^2 \propto (\eps_i^2-\eps_j^2)/\chi_{\rm Q}^2$. As we assume $\eps_i/\chi_{\rm Q}$ is a small parameter, then $(\eps_i^2-\eps_j^2)/\chi_{\rm Q}^2$ is as small or smaller, so the assumption that $\delta \ll 1$ is valid.

From equation (\ref{eqn:detprobO}) we see that the error in the output state scales as $(\eps_i^2-\eps_j^2)/\chi_{\rm Q}^2$ and is damped by the factor $\abs{\al}^4e^{-2\abs{\al}^2}/4$. For $\abs{\al} = 3$ as used in the main text this damping factor is on the order of $10^{-6}$, and so any error in the output state caused by detection probability mismatch is negligible.

A better approximation to the detection probability in the case where the JPM can also relax inelastically is \cite{Govia:2014fk}
\be
P_{i} \approx 1 - \exp\left\{-\abs{\al_i}^2\frac{\gamma_{\rm J}}{\gamma_{\rm J}+\gamma_{\rm R}}\right\}.
\ee
Using $\gamma_{\rm J} = \gamma_{\rm R}$ as elsewhere, then the error of equation (\ref{eqn:detprobO}) is damped by a factor $\propto \abs{\al}^4e^{-\abs{\al}^2}$. For $\abs{\al} = 3$ as before, this is on the order of $10^{-2}$, which again results in a negligible effect that will only decrease with increasing cavity photon number.

\section{Qubit decoherence and cavity decay}
\label{sec:decay}

When $\chi_{\rm Q}$ mismatch is present there can be a phase difference between cavity states corresponding to qubit states with the same parity, as discussed for two qubits in section \ref{sec:2Q}. In this section we examine the effect this has on intra-subspace coherence when a cavity decay mechanism is introduced, as described in the main text.

We start with a typical post drive or imperfect reset stage state of the form
\be
\label{eqn:Post}
\hat{\rho}(0) = \sum_{s,s'} c(s,s') \ket{s}\bra{s'} \otimes \hat{\rho}_{s,s'}(0) \otimes \hat{\rho}_{\rm B},
\ee
where $s, s'$ label the state of the qubits, $\hat{\rho}_{s,s'}(0)$ are the qubit dependent cavity matrices, and the environmental bath is represented by the state $\hat{\rho}_{\rm B}$, which is assumed to be stationary and a thermal state. Additionally we assume that the qubit states have no intrinsic time dependence, i.e. we work in the rotating frame of the qubits.

We consider the intial state of the qubits to be an equal superposition of two states in the same parity subspace, which for simplicity we have labelled as $\ket{0}$ and $\ket{1}$. This results in the initial system state
\begin{align}
\nonumber \hat{\rho}_{\rm S}(0) = \frac{1}{2}\Big( \ket{0}\bra{0} \otimes \ket{\al_0}\bra{\al_0} + \ket{0}\bra{1} \otimes \ket{\al_0}\bra{\al_1} \\ + \ket{1}\bra{0} \otimes \ket{\al_1}\bra{\al_0} + \ket{1}\bra{1} \otimes \ket{\al_1}\bra{\al_1}\Big),
\end{align}
where $\hat{\rho}_{s,s'}(0) = \ket{\al_s}\bra{\al_{s'}}$. Due to $\chi_{\rm Q}$ mismatch $\al_0$ and $\al_1$ differ in their phase only and we can set
\begin{align}
\al_0 = \abs{\tilde{\al}}e^{i\varphi_0}, \ \al_1 = \abs{\tilde{\al}}e^{i\varphi_1}.
\end{align}
Assuming that the cavity is in a coherent state is accurate for a post drive stage state, but the post reset state will be a more general state of the form $\hat{\rho}_{s,s'}(0) = \ket{\psi_{s}}\bra{\psi_{s'}}$ . Nevertheless, we are eventually interested in steady state dynamics which are independent of the initial state of the cavity (under the Born-Markov approximation). We will treat the post drive stage situation first and then generalize to the post reset state.

To solve for $\hat{\rho}_{s,s'}(t)$, we will use the Wigner characteristic function approach. We define the Wigner characteristic function $\chi_{ss'}(\al,t)$ of the state $\hat{\rho}_{s,s'}(t)$ as
\be
\hat{\rho}_{s,s'}(t) = \frac{1}{\pi}\int d^2\al \chi_{ss'}(\al,t)\hat{D}(-\al),
\ee
where $\hat{D}(\beta)$ is the displacement operator.

If we assume a Jaynes-Cumming type interaction between the cavity and a bosonic environment with smooth spectral density, then following \cite{Serban:2008fk} we arrive at the analytic solution for $\chi_{ss'}(\al,t)$
\be
\label{eqn:CharSol}
\chi_{ss'}(\al,t) = \chi_{ss'}(\al e^{-t(\kappa-i\omega)},0)e^{\left(\frac{\eta}{2}\abs{\al}^2(e^{-2t\kappa}-1)\right)},
\ee
where $\kappa$ is the decay rate of the cavity, $\omega$ is the frequency of the cavity mode, and $\eta = 1 + 2n(\omega)$ with $n(\omega)$ the Bose distribution at frequency $\omega$.

In order to quantify the absolute maximum amount of decoherence, we are interested in the steady state solution  ($t \rightarrow \infty$) of equation (\ref{eqn:CharSol})
\be
\label{eqn:St} 
\tilde{\chi}_{ss'}(\al) = \chi_{ss'}(0,0)e^{-\frac{\eta}{2}\abs{\al}^2}.
\ee
Using the facts that
\begin{align}
&\chi_{ss'}(0,0) = \frac{1}{4\pi}{\rm Tr}[\hat{\rho}_{s,s'}(0)]
\\ &\chi_{10}(\al,t) = \bar{\chi}_{01}(-\al,t)
\end{align}
we obtain
\begin{align}
\label{eqn:St00}
&\tilde{\chi}_{00}(\al) =  \tilde{\chi}_{11}(\al) = \frac{1}{4\pi}e^{-\frac{\eta}{2}\abs{\al}^2},
\\& \tilde{\chi}_{01}(\al)  =  {\rm exp}\left( \abs{\tilde{\al}}^2(e^{i(\varphi_1 - \varphi_0)}-1) \right) \frac{1}{4\pi}e^{-\frac{\eta}{2}\abs{\al}^2}, \label{eqn:St01}
\\& \tilde{\chi}_{10}(\al)  =  {\rm exp}\left( \abs{\tilde{\al}}^2(e^{i(\varphi_0 - \varphi_1)}-1) \right) \frac{1}{4\pi}e^{-\frac{\eta}{2}\abs{\al}^2}. \label{eqn:St10}
\end{align}
Each of equations (\ref{eqn:St00}), (\ref{eqn:St01}) and (\ref{eqn:St10}) can be reduced to $\tilde{\chi}_{ss'}(\al) = F_{ss'}(\al_0,\al_1)\chi_{\rm Thermal}(\al,\omega)$, where $\chi_{\rm Thermal}(\al,\omega) = e^{-\frac{\eta}{2}\abs{\al}^2}/4\pi$ is the Wigner characteristic function for a thermal state at frequency $\omega$.

In light of this simplification the steady state solution for the full system reduces to
\begin{align}
\nonumber&\hat{\rho}_S = \sum_{s,s'} c(s,s')F_{ss'}(\al_0,\al_1) \ket{s}\bra{s} 
\\ &\otimes\frac{1}{\pi} \int d^2{\al}\chi_{\rm Thermal}(\al,\omega)\hat{D}(-\al),
\end{align}
where $F_{ss'}(\al_s,\al_{s'}) =  {\rm exp}\left( \abs{\tilde{\al}}^2(e^{i(\varphi_s - \varphi_{s'})}-1)\right)$. As this is a product state it is the trivial to trace out the state of the cavity and obtain the state of the qubits alone
\begin{align}
\nonumber&\hat{\rho}_Q = \frac{1}{2}\Big( \ket{0}\bra{0} + F_{01}(\al_0,\al_1)\ket{0}\bra{1} 
\\&+ F_{10}(\al_1,\al_0)\ket{1}\bra{0} + \ket{1}\bra{1} \Big).
\end{align}

For two qubits, we examine the damping envelope of the off diagonal elements, given by
\be
F_{01}(\al_0,\al_1) = \exp{\left( -\abs{\tilde{\al}}^2(e^{2i\varphi_0}+1) \right)},
\ee
where we have used the fact that $\varphi_0 - \varphi_1 = 2\varphi_0 - \pi$ for two qubits, as can be calculated from either pair of expressions in equation (\ref{eqn:als}). We are interested in the absolute value of this envelope, which is given by
\begin{align}
\nonumber \abs{F_{01}(\al_0,\al_1)} &= \exp{\left( -\abs{\tilde{\al}}^2{\rm Re}\left\{(e^{2i\varphi_0}+1) \right\}\right)} \\
& = \exp{\left[ -\abs{\tilde{\al}}^2\left(1 - \cos{\left(\frac{\pi}{\chi_{\rm Q}}\eps\right)}\right)\right]}, \label{eqn:DampM}
\end{align}
where the second equality comes from using equation (\ref{eqn:als}) to define $e^{2i\varphi_0}$. From this we see that the decoherence depends on both the cavity occupation and the magnitude of the $\chi_{\rm Q}$ mismatch, being maximal when $\eps/\chi_{\rm Q} = 1$. Using $\abs{\al}^2 = 9$ as used elsewhere, and a strong mismatch of $\eps/\chi_{\rm Q} = 0.1$, the qubit coherence is reduced to 64\% in the steady state. However, for a high-Q cavity we do not expect any photon loss during the parity readout protocol, and as such, the post reset state is of greater interest.

Generalizing to the post reset state we need only change the definition of the damping envelope to
\be
\label{eqn:Damp}
F_{ss'} = {\rm Tr}[\hat{\rho}_{s,s'}(0)] = \braket{\psi_{s'}}{\psi_s},
\ee
where $\ket{\psi_s}$ is the state after detection and imperfect reset for an initial cavity state $\ket{\al_s}$. We will consider the worst case scenario for the states $\ket{\al_{0/1}}$, where the damping envelope of equation (\ref{eqn:DampM}) reduces to $\exp(-2\abs{\tilde{\al}}^2)$. This occurs for $\eps = \chi_{\rm Q}$, where the states $\ket{\al_{0/1}}$ lie on the real axis with $\al_0 = -\al_1$. This is the worst case as the states are maximally distinguishable, and correspondingly we expect the states $\ket{\psi_{0/1}}$ to also be maximally distinguishable in this case.

\begin{figure}[h!]
\includegraphics[width = \columnwidth]{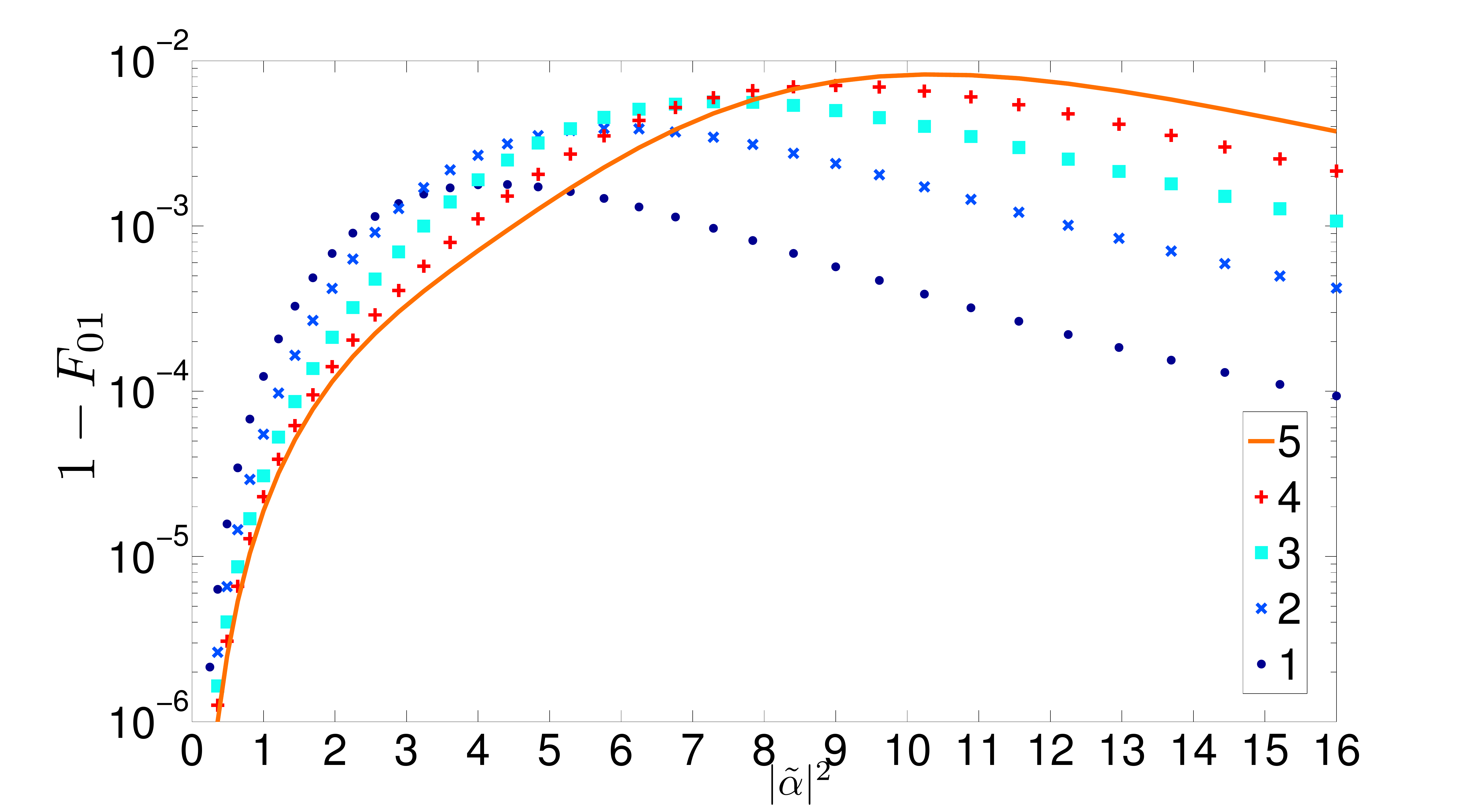}
\caption{$1-F_{01}$ as a function of the cavity occupation prior to detection and imperfect reset (horizontal axis), and the number of photons removed from the cavity by the photon detector (color and line style). The initial cavity states considered are maximally distinguishable, and so this represents the worst case scenario.}
\label{fig:Damp}
\end{figure}

Figure \ref{fig:Damp} shows a numerical simulation of $1-F_{01}$ for the worst case post reset state as a function of both the initial cavity occupation $\abs{\tilde{\al}}^2$, and the number of photons removed by the photon detector. For these simulations, the detector was assumed to be a JPM with subtraction operator back action \cite{Govia2014}. Multiple photons can be removed due to JPM inelastic relaxation. As can be seen, even for the worst case scenario, qubit coherence is reduced by less than $1\%$. As we expect $\eps \ll \chi_{\rm Q}$ in a realistic experiment, the decoherence caused by post reset state photon loss will be far from the upper bound presented here, and can be considered inconsequential.

\section{Higher Order Effects Beyond the Dispersive Hamiltonian}
\label{sec:SHigh}

While the dispersive Hamiltonian is the first order approximation to the cavity-qubit coupling in the relevant regime of our protocol, it is worthwhile to consider the effects on the protocol of the full Jaynes-Cummings Hamiltonian for the cavity-qubit coupling, given by
\begin{equation}
\hat{H}_{\rm JC} = \omega_{\rm C} \hat a ^\dagger \hat a -\sum_{k=0}^{N}\frac{1}{2}\omega_{\rm Q} \hat \sigma_z^k + \sum_{k=0}^{N}g_{k}\left(\hat\sigma_{+}^k\hat{a} + \hat\sigma_{-}^k\hat{a}^{\dagger}\right),
\label{eq:HJC}
\end{equation}
where $g_{k}$ is the cavity-qubit coupling for qubit $k$. The major effects of the full Jaynes-Cummings Hamiltonian occur during the drive stage. 

\begin{figure}[h!]
\subfigure{
\label{fig:4QDriveJC}
\includegraphics[width = \columnwidth]{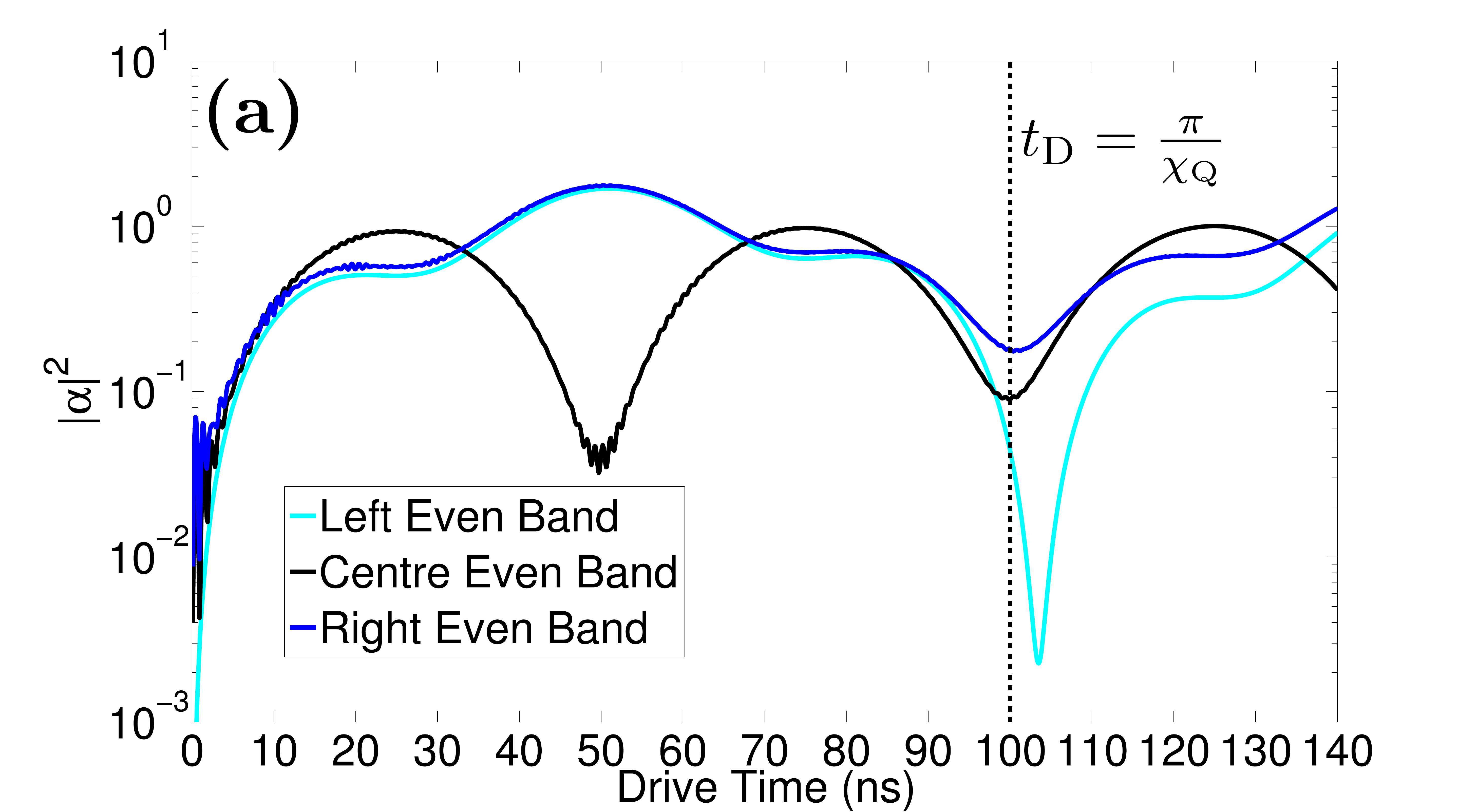}}
\subfigure{
\label{fig:4QMeasJC}
\includegraphics[width = \columnwidth]{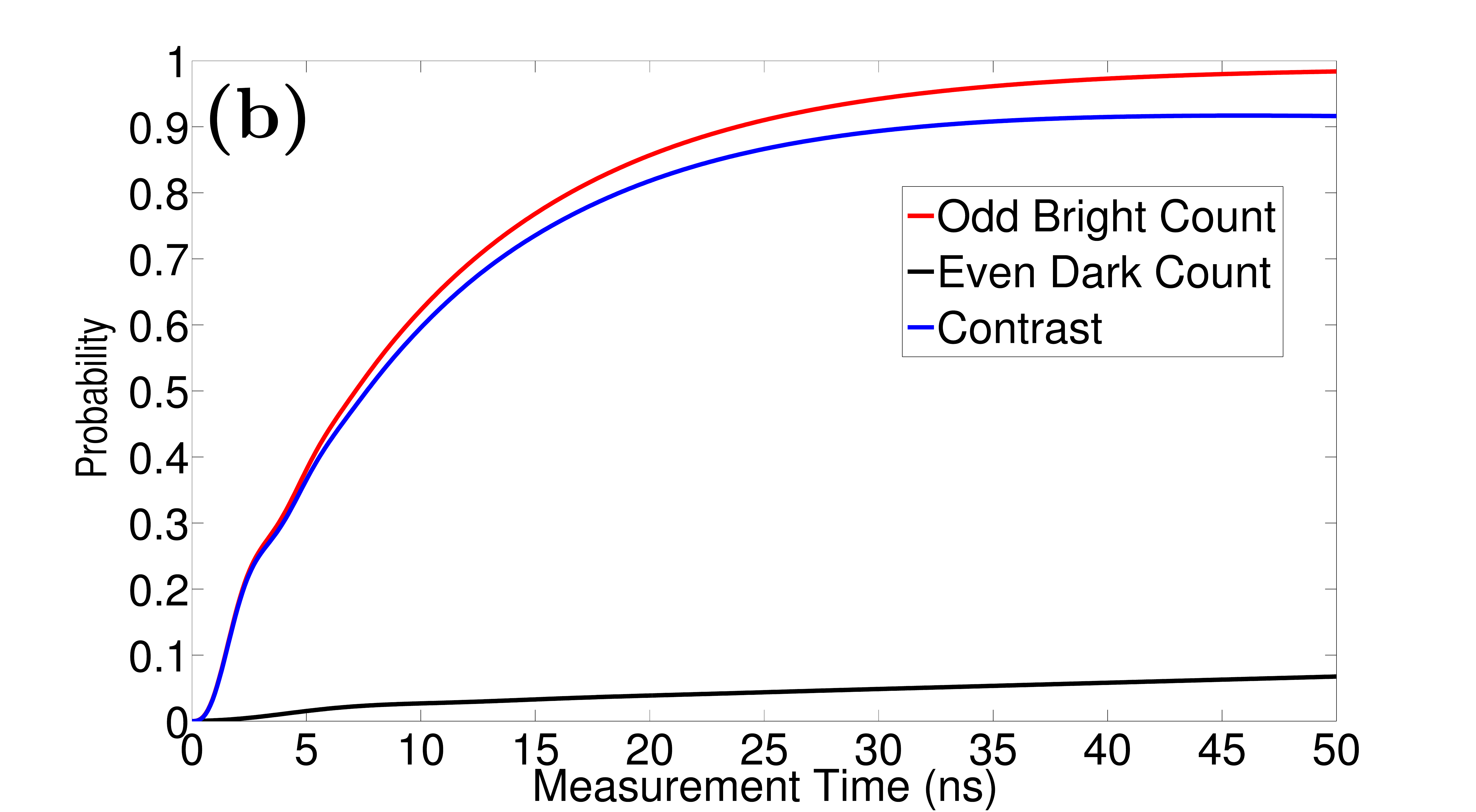}}
\caption{{\bf(a)} Cavity occupation for even-parity qubit states with Jaynes-Cummings cavity-qubit coupling. {\bf (b)} Bright and dark count rates, and measurement contrast for the cavity photon numbers of {\bf(a)}, assuming the even state is from the right band (worst case).}
\end{figure}

The first is an asymmetric shift to the cavity frequency within a parity subspace, due to the Kerr-like interaction term, which is one order of approximation higher than the dispersive Hamiltonian. The result of this is that the minima of the even qubit parity cavity occupation curves no longer coincide, as can be seen in FIG. \ref{fig:4QDriveJC} for the four qubit case. The second major effect is the formation of dressed cavity-qubit states, the result of which is that there is residual cavity occupation for even-parity states, and the minima are no longer zero for all even-parity states. Therefore, for no choice of $t_{\rm D}$ will the cavity be in the vacuum state if the qubits are in an even-parity state. This reduces the measurement contrast, though as can be seen in figure FIG. \ref{fig:4QMeasJC} this reduction is only on the order of a few percent, and measurement contrasts approaching 92\% can still be achieved with the same parameters as in the main text.

These effects can be mitigated by increasing the cavity-qubit detuning, in which case the parity measurement contrast asymptotes to the dispersive value. However, this either decreases $\chi_{\rm Q}$ and therefore increases the length of the drive stage, or necessitates also increasing the cavity-qubit couplings (to keep $\chi_{\rm Q}$ constant), which is only advisable to the point where the rotating wave approximation begins to break down. It is also possible to improve the contrast by better pulse shaping in the drive stage, and this will be the focus of future work. As can also be seen in FIG. \ref{fig:4QDriveJC}, the fact that the even qubit parity cavity photon states are not the same between parity bands would cause intra-subspace qubit decoherence. Full coherence would be returned by perfect reset, unless photon loss occurs (either through a detection or other mechanism). However, photon loss of any kind is so unlikely for these low occupation cavity states that the effect on qubit coherence is minimal, on the order of a few percent at most.

\end{document}